7.3.20

# Solution of the Monoenergetic Neutron Transport Equation in a Half Space

B. Ganapol
Department of Aerospace and Mechanical Engineering
University of Arizona

The analytical solution of neutron transport equation has fascinated mathematicians and physicists alike since the Milne half-space problem was introduce in 1921 [1]. Numerous numerical solutions exist, but understandably, there are only a few analytical solutions, with the prominent one being the singular eigenfunction expansion (SEE) introduced by Case [2] in 1960. For the half-space, the method, though yielding, an elegant analytical form resulting from half-range completeness, requires numerical evaluation of complicated integrals. In addition, one finds closed form analytical expressions only for the infinite medium and half-space cases. One can find the flux in a slab only iteratively. That is to say, in general one must expend a considerable numerical effort to get highly precise benchmarks from SEE. As a result, investigators have devised alternative methods, such as the CN [3], FN [4] and Greens Function Method (GFM) [5] based on the SEE have been devised. These methods take the SEE at their core and construct a numerical method around the analytical form. The FN method in particular has been most successful in generating highly precise benchmarks. No method yielding a precise numerical solution has yet been based solely on a fundamental discretization until now. Here, we show for the albedo problem with a source on the vacuum boundary of a homogeneous medium, a precise numerical solution is possible via Lagrange interpolation over a discrete set of directions.

Since this is an initial progress report of a new solution, we will consider only the simplest case in the half-space. In particular, the source will be isotropic and the medium isotropically scattering.

## 1. General solution to 1D transport equation in a half-space with source

We begin with the 1D transport equation for general anisotropic scattering

$$\left[\mu\frac{\partial}{\partial x}+1\right]\psi(x,\mu) = \frac{c}{2}\int_{-1}^{1}d\mu' f(\mu,\mu')\psi(x,\mu') + S(x,\mu), \tag{1}$$

with a general volume source and subject to an incoming source at the free surface $\psi(0,\mu), 0 \leq \mu \leq 1$. One can represent the solution by a combination of the solutions to the homogeneous form

$$\left[\mu\frac{\partial}{\partial x}+1\right]\psi_h(x,\mu) = \frac{c}{2}\int_{-1}^{1} d\mu' f(\mu,\mu')\psi_h(x,\mu') \tag{2a}$$

and *any* particular solution including the source

$$\left[\mu\frac{\partial}{\partial x}+1\right]\psi_p(x,\mu) = \frac{c}{2}\int_{-1}^{1} d\mu' f(\mu,\mu')\psi_p(x,\mu') + S(x,\mu) \tag{2b}$$

as

$$\psi(x,\mu) = \psi_h(x,\mu) + \psi_p(x,\mu). \tag{2c}$$

The boundary condition then applies to the combination.

The particular solution is therefore

$$\psi_p(x,\mu) = \int_{-\infty}^{\infty} dx' \int_{-1}^{1} d\mu' G(x-x',\mu,\mu') S(x',\mu'), \tag{3a}$$

where the Green's function [2] is for $x > x'$

$$G_>(x-x',\mu,\mu') = \phi_{0+}(\mu')\phi_{0+}(\mu)\frac{e^{-|x-x'|/v_0}}{N_{0+}} + \int_0^1 dv' \frac{e^{-|x-x'|/v'}}{N_{v'}} \phi_{v'}(\mu')\phi_{v'}(\mu) \tag{3b}$$

and for $x < x'$

$$G_<(x-x',\mu,\mu') = -\phi_{0-}(\mu')\phi_{0-}(\mu)\frac{e^{-|x-x'|/v_0}}{N_{0-}} + \int_0^1 dv' \frac{e^{-|x-x'|/v'}}{N_{-v'}} \phi_{-v'}(\mu')\phi_{-v'}(\mu). \tag{3c}$$

Starting from the SEE and with considerable algebra, we now derive an expression for the general solution noting

$$N_{0\pm} \equiv \frac{c}{2} v_0^3 \left[ \frac{c}{v_0^2 - 1} - \frac{1}{v_0^2} \right]$$

$$N_v \equiv v\left[ (1 - cv \tanh^{-1} v)^2 + (c\pi v/2)^2 \right]$$

(3d,e)

in the equations above.

The solution to the homogeneous equation by SEE is

$$\psi_h(x,\mu) = a_{0+}\phi_{0+}(\mu)e^{-x/v_0} + a_{0-}\phi_0(\mu)e^{-x/v_0} + \int_{-1}^{1} dv' e^{-x/v'} \phi_{v'}(\mu) A(v') \quad (4)$$

giving the general solution from Eqs(3) and (4)

$$\psi(x,\mu) = a_{0+}\phi_{0+}(\mu)e^{-x/v_0} + a_{0-}\phi_0(\mu)e^{x/v_0} + \int_{-1}^{1} dv' e^{-x/v'} \phi_{v'}(\mu) A(v')$$

$$+ \int_{-\infty}^{x} dx' \int_{-1}^{1} d\mu' \left[ \phi_{0+}(\mu')\phi_{0+}(\mu) \frac{e^{-|x-x'|/v_0}}{N_{0+}} + \int_{0}^{1} dv' \frac{e^{-|x-x'|/v'}}{N_{v'}} \phi_{v'}(\mu')\phi_{v'}(\mu) \right] S(x',\mu') +$$

$$+ \int_{x}^{\infty} dx' \int_{-1}^{1} d\mu' \left[ -\phi_{0-}(\mu')\phi_{0-}(\mu) \frac{e^{-|x-x'|/v_0}}{N_{0-}} + \int_{0}^{1} dv' \frac{e^{-|x-x'|/v'}}{N_{-v'}} \phi_{-v'}(\mu')\phi_{-v'}(\mu) \right] S(x',\mu').$$

(5)

From orthogonality at $x = 0$

$$\int_{-1}^{1} d\mu\mu\phi_{0+}(\mu)\psi(0,\mu) = a_{0+}N_{0+} + \int_{-\infty}^{0} dx' \int_{-1}^{1} d\mu'\phi_{0+}(\mu')e^{-|x'|/v_0} S(x',\mu') \quad (6a)$$

$$\int_{-1}^{1} d\mu\mu\phi_{0-}(\mu)\psi(0,\mu) = a_{0-}N_{0-} - \int_{0}^{\infty} dx' \int_{-1}^{1} d\mu'\phi_{0-}(\mu')e^{-|x'|/v_0} S(x',\mu') \quad (6b)$$

$$\int_{-1}^{1} d\mu\mu\phi_{v}(\mu)\psi(0,\mu) = N_v A(v) + \int_{-\infty}^{0} dx' \int_{-1}^{1} d\mu' e^{-|x'|/v} \phi_{v}(\mu') S(x',\mu') \quad (6c)$$

$$\int_{-1}^{1} d\mu\mu\phi_{-v}(\mu)\psi(0,\mu) = N_{-v} A(-v) + \int_{0}^{\infty} dx' \int_{-1}^{1} d\mu' e^{-|x'|/v} \phi_{-v}(\mu') S(x',\mu'). \quad (6d)$$

Then solving for the expansion coefficients gives

$$a_{0+} = \frac{1}{N_{0+}}\left\{\int_{-1}^{1} d\mu\,\mu\phi_{0+}(\mu)\psi(0,\mu) - \int_{-\infty}^{0} dx'\int_{-1}^{1} d\mu'\phi_{0+}(\mu')e^{-|x'|/v_0}S(x',\mu')\right\} \quad (7a)$$

$$a_{0-} = \frac{1}{N_{0-}}\left\{\int_{-1}^{1} d\mu\,\mu\phi_{0-}(\mu)\psi(0,\mu) + \int_{0}^{\infty} dx'\int_{-1}^{1} d\mu'\phi_{0-}(\mu')e^{-|x'|/v_0}S(x',\mu')\right\} \quad (7b)$$

$$A(v) = \frac{1}{N_v}\left\{\int_{-1}^{1} d\mu\,\mu\phi_v(\mu)\psi(0,\mu) - \int_{-\infty}^{0} dx'\int_{-1}^{1} d\mu'e^{-|x'|/v}\phi_v(\mu')S(x',\mu')\right\} \quad (7c)$$

$$A(-v) = \frac{1}{N_{-v}}\left\{\int_{-1}^{1} d\mu\,\mu\phi_{-v}(\mu)\psi(0,\mu) - \int_{0}^{\infty} dx'\int_{-1}^{1} d\mu'e^{-|x'|/v}\phi_{-v}(\mu')S(x',\mu')\right\}. \quad (7d)$$

On substitution into Eq(5)

$$\psi(x,\mu) = \frac{1}{N_{0+}}\left\{\int_{-1}^{1} d\mu'\mu'\phi_{0+}(\mu')\psi(0,\mu') - \int_{-\infty}^{0} dx'\int_{-1}^{1} d\mu'\phi_{0+}(\mu')e^{x'/v_0}S(x',\mu')\right\}\phi_{0+}(\mu)e^{-x/v_0} +$$

$$+\frac{1}{N_{0-}}\left\{\int_{-1}^{1} d\mu'\mu'\phi_{0-}(\mu')\psi(0,\mu') + \int_{0}^{\infty} dx'\int_{-1}^{1} d\mu'\phi_{0-}(\mu')e^{-x'/v_0}S(x',\mu')\right\}\phi_{0-}(\mu)e^{x/v_0} +$$

$$+\int_{-\infty}^{x} dx'\int_{-1}^{1} d\mu'\phi_{0+}(\mu')\phi_{0+}(\mu)\frac{e^{-(x-x')/v_0}}{N_{0+}}S(x',\mu') +$$

$$-\int_{x}^{\infty} dx'\int_{-1}^{1} d\mu'\phi_{0-}(\mu')\phi_{0-}(\mu)\frac{e^{-(x'-x)/v_0}}{N_{0-}}S(x',\mu') +$$

$$+\int_{0}^{1} dv'\frac{1}{N_{v'}}\left\{\int_{-1}^{1} d\mu'\mu'\phi_{v'}(\mu')\psi(0,\mu') - \int_{-\infty}^{0} dx'\int_{-1}^{1} d\mu'e^{x'/v'}\phi_{v'}(\mu')S(x',\mu')\right\}e^{-x/v'}\phi_{v'}(\mu) +$$

$$+\int_{0}^{1} dv'\frac{1}{N_{-v'}}\left\{\int_{-1}^{1} d\mu'\mu'\phi_{-v'}(\mu')\psi(0,\mu') - \int_{0}^{\infty} dx'\int_{-1}^{1} d\mu'e^{-x'/v'}\phi_{-v'}(\mu')S(x',\mu')\right\}e^{x/v'}\phi_{-v'}(\mu) +$$

$$+\int_{-\infty}^{x} dx'\int_{-1}^{1} d\mu'\int_{0}^{1} dv'\frac{e^{-(x-x')/v'}}{N_{v'}}\phi_{v'}(\mu')\phi_{v'}(\mu)S(x',\mu') +$$

$$+\int_{x}^{\infty} dx'\int_{-1}^{1} d\mu'\int_{0}^{1} dv'\frac{e^{-(x'-x)/v'}}{N_{-v'}}\phi_{-v'}(\mu')\phi_{-v'}(\mu)S(x',\mu') \quad (8)$$

and on combination of terms, we finally find

$$\psi(x,\mu) = \frac{1}{N_{0+}}\left\{\int_{-1}^{1}d\mu'\left\{\mu'\psi(0,\mu') + \int_{0}^{x}dx'e^{x'/\nu_0}S(x',\mu')\right\}\phi_{0+}(\mu')\right\}\phi_{0+}(\mu)e^{-x/\nu_0} +$$

$$+ \frac{1}{N_{0-}}\left\{\int_{-1}^{1}d\mu'\left\{\mu'\psi(0,\mu') + \int_{0}^{x}dx'e^{-x'/\nu_0}S(x',\mu')\right\}\phi_{0-}(\mu')\right\}\phi_{0-}(\mu)e^{x/\nu_0} +$$

$$+ \int_{-1}^{1}d\nu'\frac{1}{N_{\nu'}}\left\{\int_{-1}^{1}d\mu'\left\{\mu'\psi(0,\mu') + \int_{0}^{x}dx'e^{x'/\nu'}S(x',\mu')\right\}\phi_{\nu'}(\mu')\right\}e^{-x/\nu'}\phi_{\nu'}(\mu)$$

(9)

to give on re-arrangement

$$\psi(x,\mu) = \alpha_{0+}\phi_{0+}(\mu)e^{-x/\nu_0} + \alpha_{0-}\phi_{0-}(\mu)e^{x/\nu_0} + \int_{-1}^{1}d\nu'e^{-x/\nu'}\phi_{\nu'}(\mu)A(\nu') + \psi_P(x,\mu)$$

(10a)

$$\psi_P(x,\mu) \equiv \frac{1}{N_{0+}}\int_{-1}^{1}d\mu'\int_{0}^{x}dx'e^{x'/\nu_0}S(x',\mu')\phi_{0+}(\mu')\phi_{0+}(\mu)e^{-x/\nu_0} +$$

$$+ \frac{1}{N_{0-}}\int_{-1}^{1}d\mu'\int_{0}^{x}dx'e^{-x'/\nu_0}S(x',\mu')\phi_{0-}(\mu')\phi_{0-}(\mu)e^{x/\nu_0} + \quad (10b)$$

$$+ \int_{-1}^{1}d\nu'\frac{e^{-x/\nu'}}{N_{\nu'}}\left[\int_{-1}^{1}d\mu\int_{0}^{x}dx'e^{x'/\nu'}S(x',\mu')\phi_{\nu'}(\mu')\right]\phi_{\nu'}(\mu),$$

where

$$\alpha_{0\pm} \equiv \frac{1}{N_{0\pm}}\int_{-1}^{1}d\mu\mu\phi_{0\pm}(\mu)\psi(0,\mu) \tag{10c}$$

$$A(\nu) \equiv \frac{1}{N_\nu}\int_{-1}^{1}d\mu\mu\phi_\nu(\mu)\psi(0,\mu). \tag{10d}$$

For a half-space, the flux must also be finite at infinity, so

$$\lim_{x\to\infty}\psi(x,\mu)<\infty, \tag{11}$$

which from Eqs(10) says

$$\begin{aligned}&\alpha_{0-}\equiv 0\\&A(v)\equiv 0;\ -1\leq v\leq 0\\&\lim_{x\to\infty}\psi_P(x,\mu)<\infty.\end{aligned} \tag{12a,b,c}$$

This seems to indicate from the first term in Eq(10b)

$$\lim_{x\to\infty}\left[e^{-x/v_0}\int_0^x dx'e^{x'/v_0}S(x',\mu')\right]=\lim_{x\to\infty}\left[\frac{\int_0^x dx'e^{x'/v_0}S(x',\mu')}{e^{x/v_0}}\right] \tag{13a}$$

$$=v_0\lim_{x\to\infty}S(x,\mu')<\infty$$

and from the second term

$$\lim_{x\to\infty}\left[e^{x/v_0}\int_0^x dx'e^{-x'/v_0}S(x',\mu')\right]=\lim_{x\to\infty}\left[\frac{\int_0^x dx'e^{-x'/v_0}S(x',\mu')}{e^{-x/v_0}}\right] \tag{13b}$$

$$=v_0\lim_{x\to\infty}S(x,\mu')<\infty$$

and similarly from the third term for $-1\leq v\leq 0$.

## 2. Lagrange interpolation
Consider the solution to the following albedo problem

$$\psi(x,\mu)=a_{0+}\phi_{0+}(\mu)e^{-x/v_0}+\int_0^1 dve^{-x/v}\phi_v(\mu)A(v), \tag{14}$$

where the incoming flux at the free surface is $\psi(0,\mu)$, $0 \leq \mu \leq 1$. Note that one constructs solution such that the solution vanishes at infinity. Consequently

$$\psi(0,-\mu) = a_{0+}\phi_{0+}(-\mu) + \int_0^1 dv\,\phi_v(-\mu)A(v) \tag{15a}$$

$$\psi(0,\mu) = a_{0+}\phi_{0+}(\mu) + \int_0^1 dv\,\phi_v(\mu)A(v), \tag{15b}$$

where

$$\phi_{0+}(\mu) = \frac{cv_0}{2}\frac{1}{v_0-\mu}$$

$$\phi_v(\mu) = \frac{cv}{2}P\frac{1}{v-\mu} + \lambda(v)\delta(v-\mu). \tag{15c,d}$$

Thus

$$\lambda(v)A(v) + \frac{c}{2}\int_{-1}^{1} dv'\frac{v'}{v'-v}A(v') = \psi(0,v) - a_{0+}\frac{cv_0}{2}\frac{1}{v_0-v}. \tag{16}$$

If

$$A(v) = A_1(v) - a_{0+}A_2(v), \tag{17a}$$

then

$$\lambda(v)A_1(v) + \frac{c}{2}\int_{-1}^{1} dv'\frac{v'}{v'-v}A_1(v') = \psi(0,v)$$

$$\lambda(v)A_2(v) + \frac{c}{2}\int_{-1}^{1} dv'\frac{v'}{v'-v}A_2(v') = \frac{cv_0}{2}\frac{1}{v_0-v}, \tag{17b,c}$$

which gives

$$\psi(0,-\mu) = a_{0+}\left[\phi_{0+}(-\mu) - \int_0^1 d\nu \phi_\nu(-\mu) A_2(\nu)\right] + \int_0^1 d\nu \phi_\nu(-\mu) A_1(\nu) \quad (18a)$$

with

$$a_{0+} = \frac{1}{N_{0+}} \int_{-1}^{1} d\mu\mu\phi_{0+}(\mu)\psi(0,\mu)$$
$$= \frac{1}{N_{0+}}\left[\int_0^1 d\mu\mu\phi_{0+}(\mu)\psi(0,\mu) - \int_0^1 d\mu\mu\phi_{0+}(-\mu)\psi(0,-\mu)\right]. \quad (18b)$$

Introducing $\psi(0,-\mu)$ from Eq(18a) gives

$$a_{0+} = \frac{1}{N_{0+}}\left\{ \begin{array}{l} \int_0^1 d\mu\mu\phi_{0+}(\mu)\psi(0,\mu) - \\ -a_{0+}\left[J_{0+} - \int_0^1 d\nu J_\nu A_2(\nu)\right] - \int_0^1 d\nu J_\nu A_1(\nu) \end{array} \right\} \quad (19a)$$

with

$$J_\nu = \int_0^1 d\mu\mu\phi_{0+}(-\mu)\phi_\nu(-\mu) \quad (19b)$$

or

$$J_\nu = \begin{cases} \left(\dfrac{c\nu}{2}\right)\phi_{0+}(\nu)\left[\nu_0 \ln\left[\dfrac{\nu_0+1}{\nu_0}\right] - \nu \ln\left[\dfrac{\nu+1}{\nu}\right]\right], & 0 \leq \nu \leq 1 \\ \left(\dfrac{c\nu_0}{2}\right)\left[\ln\left[\dfrac{\nu_0+1}{\nu_0}\right] - \dfrac{1}{\nu_0+1}\right], & \nu = 0+. \end{cases} \quad (19c)$$

Solving for $a_{0+}$ in Eq(19a)

$$a_{0+} = \left\{ N_{0+} + \left[ J_{0+} - \int_0^1 dv J_v A_2(v) \right] \right\}^{=1} \left\{ \int_0^1 d\mu\mu\phi_{0+}(\mu)\psi(0,\mu) - \int_0^1 dv J_v A_1(v) \right\}$$
(20)

then provides $a_{0+}$ in terms of the solution to the singular integral equations of Eqs(17bc) to be numerically solved in the next section.

## 3. Numerical considerations

Consider Eqs(17bc) as

$$\lambda(v)A_i(v) + \frac{c}{2}\int_{-1}^1 dv' \frac{v'}{v'-v} A_i(v') = f_i, \tag{21a}$$

where

$$f_i(v) = \begin{cases} \psi(0,v), & i=1 \\ \phi_{0+}(v), & i=2. \end{cases} \tag{21b,c}$$

The desired solution is in terms of discrete values of the variable $A_i(v)$. On discretization of Eqs(21), one must use caution because of the principal value integration. The principal value integral is most efficiently accommodated however through Lagrange interpolation

$$A_i(v) = \sum_{j=1}^N l_j(v) A_{ij} \tag{22a}$$

with

$$l_j(v) = \frac{P_N^*(v)}{(v-v_j)P_N^{*'}(v_j)} \tag{22b}$$

and $v$ are the zeros of the polynomial $P_N^*(v) = 0$; $v = v_j$, $j = 1,...,N$. WE consider only half-range Legendre polynomials where $P_N^*(v) \equiv P_N(2v-1)$.

When introduced into Eq(21a), there results

$$\lambda(v) A_i(v) + \sum_{j=1}^{N} I_j(v) A_{ij} = f_i(v) \qquad (23a)$$

with

$$I_j(v) \equiv \frac{c}{2} \int_0^1 dv' \frac{v'}{(v'-v)} l_j(v'). \qquad (23b)$$

or with Eq(22b)

$$I_j(v) \equiv \frac{c}{2 P_N^{*'}(v_j)} \int_0^1 dv' \left[ \frac{v'}{(v'-v)} \frac{P_N^*(v')}{(v'-v_j)} \right]. \qquad (23c)$$

The integral conveniently evaluates to

$$I_j(v) = -\frac{c}{2 P_N^{*'}(v_j)} \begin{cases} \frac{2}{v_j - v} \left[ v_j Q_N(2v_j - 1) - v Q_N(2v - 1) \right], & v \neq v_j \\ 2 Q_N(2v_j - 1), & v = v_j. \end{cases}$$

$$(24)$$

When evaluated at the zero $v_m, m = 1,...,N$, Eq(23a) becomes

$$\sum_{j=1}^{N} \left\{ \left[ \lambda(v_m) + I_m(v_m) \right] \delta_{jm} + (1 - \delta_{jm}) I_j(v_m) \right\} A_{ij} = f_{im}, \quad m = 1,...,N.$$

$$(25)$$

Thus, the analytical form for $a_{0+}$ is

$$a_{0+} = \left\{ N_{0+} + \left[ J_{0+} - K_1 \right] \right\}^{-1} \left\{ \int_0^1 d\mu \mu \phi_{0+}(\mu) \psi(0,\mu) - K_2 \right\}, \qquad (26a)$$

where

$$K_i \equiv K_{i1} + K_{i2} \tag{26b}$$

and

$$K_{i1} \equiv \frac{cv_0}{2} \ln\left[\frac{v_0+1}{v_0}\right] \int_0^1 dv v \phi_{0+}(v) A_i(v)$$

$$K_{i2} \equiv \frac{cv_0}{2} \int_0^1 dv v^2 \phi_{0+}(v) \ln\left[\frac{v+1}{v}\right] A_i(v) \tag{26c,d}$$

Seems that $K_{i1}$ can be identified as

$$K_{i1} \equiv \left(\frac{cv_0}{2}\right)^2 \ln\left[\frac{v_0+1}{v_0}\right] \sum_{j=1}^N A_{ij} \int_0^1 dv \frac{v}{v_0-v} l_j(v)$$

to give

$$K_{i1} \equiv -\left(\frac{cv_0}{2}\right)^2 \ln\left[\frac{v_0+1}{v_0}\right] \sum_{j=1}^N I_j(v_0) A_{ij}$$

$$K_{i1} \equiv 2\left(\frac{cv_0}{2}\right)^2 \ln\left[\frac{v_0+1}{v_0}\right] \sum_{j=1}^N \frac{A_{ij}}{P_N^{*'}(v_j)} \frac{1}{v_j-v_0}\left[v_j Q_N(2v_j-1) + v_0 Q_N(2v_0-1)\right].$$

$$\tag{27}$$

$K_{i2}$ has defied analytical efforts to evaluate and is perform via Gauss quadrature.

Once $a_{0+}$ and $A_{ij}, i=1,2; j=1,...,N$ are known, Eq(15a) gives the exiting flux as

$$\psi(0,-\mu) = a_{0+}\left[\phi_{0+}(-\mu) - \sum_{j=1}^N I_j(-\mu) A_{2j}\right] + \int_0^1 dv \sum_{j=1}^N I_j(-\mu) A_{1j}, \tag{28}$$

with substitution of Eq(23b).

## 4. Numerical Results

We now consider a uniform source striking the free surface of a semi-infinite isotopically scattering homogeneous medium. We use iteration in $N$ to find the exiting flux at 10 uniformly spaced directions to at least 7-places. The iteration advances in $N$ with a stride of four until convergence. Wynn-epsilon (W-e) acceleration [6] enhances convergence. All digits shown in Table 1 are identical in comparison to the response matrix method [7]. Table 2 gives an addition benchmark where the precision is expected to be better than one unit in the last place.

Table 1 Seven place benchmark for albedo problem for $0.4 \leq c \leq 0.9$

| $\mu \backslash c$ | 0.4 | 0.5 | 0.6 | 0.7 | 0.8 | 0.9 |
|---|---|---|---|---|---|---|
| 1.0000E+00 | 8.3357277E-02 | 1.1522588E-01 | 1.5541466E-01 | 2.0867995E-01 | 2.8525450E-01 | 4.1494748E-01 |
| 9.0000E-01 | 8.8446308E-02 | 1.2198994E-01 | 1.6408263E-01 | 2.1951910E-01 | 2.9852634E-01 | 4.3054104E-01 |
| 8.0000E-01 | 9.4223842E-02 | 1.2963262E-01 | 1.7381852E-01 | 2.3159952E-01 | 3.1315688E-01 | 4.4742308E-01 |
| 7.0000E-01 | 1.0084825E-01 | 1.3834805E-01 | 1.8484644E-01 | 2.4516406E-01 | 3.2938544E-01 | 4.6578138E-01 |
| 6.0000E-01 | 1.0853464E-01 | 1.4839744E-01 | 1.9746445E-01 | 2.6053111E-01 | 3.4752006E-01 | 4.8585154E-01 |
| 5.0000E-01 | 1.1758562E-01 | 1.6014443E-01 | 2.1208237E-01 | 2.7813176E-01 | 3.6796976E-01 | 5.0793891E-01 |
| 4.0000E-01 | 1.2844760E-01 | 1.7411918E-01 | 2.2928957E-01 | 2.9857597E-01 | 3.9130218E-01 | 5.3245844E-01 |
| 3.0000E-01 | 1.4182526E-01 | 1.9114797E-01 | 2.4999130E-01 | 3.2278524E-01 | 4.1835924E-01 | 5.6001640E-01 |
| 2.0000E-01 | 1.5895774E-01 | 2.1266387E-01 | 2.7573436E-01 | 3.5230943E-01 | 4.5053604E-01 | 5.9161280E-01 |
| 1.0000E-01 | 1.8252201E-01 | 2.4172077E-01 | 3.0977054E-01 | 3.9036735E-01 | 4.9070973E-01 | 6.2933582E-01 |
| 0.0000E+00 | 2.2540333E-01 | 2.9289321E-01 | 3.6754446E-01 | 4.5227744E-01 | 5.5278640E-01 | 6.8377223E-01 |

Table 1 Seven place benchmark for albedo problem for $0.99 \leq c \leq 0.99999$

| $\mu \backslash c$ | 0.99 | 0.999 | 0.9999 | 0.99999 |
|---|---|---|---|---|
| 1.0000E+00 | 7.5272072E-01 | 9.1284533E-01 | 9.7141778E-01 | 9.9085481E-01 |
| 9.0000E-01 | 7.6430578E-01 | 9.1772948E-01 | 9.7311399E-01 | 9.9140750E-01 |
| 8.0000E-01 | 7.7629964E-01 | 9.2268485E-01 | 9.7482251E-01 | 9.9196288E-01 |
| 7.0000E-01 | 7.8873759E-01 | 9.2771823E-01 | 9.7654524E-01 | 9.9252152E-01 |
| 6.0000E-01 | 8.0166441E-01 | 9.3283900E-01 | 9.7828488E-01 | 9.9308429E-01 |
| 5.0000E-01 | 8.1513990E-01 | 9.3806086E-01 | 9.8004549E-01 | 9.9365243E-01 |
| 4.0000E-01 | 8.2924965E-01 | 9.4340513E-01 | 9.8183349E-01 | 9.9422798E-01 |
| 3.0000E-01 | 8.4412871E-01 | 9.4890800E-01 | 9.8365999E-01 | 9.9481442E-01 |
| 2.0000E-01 | 8.6002328E-01 | 9.5463979E-01 | 9.8554678E-01 | 9.9541861E-01 |
| 1.0000E-01 | 8.7751247E-01 | 9.6077432E-01 | 9.8754820E-01 | 9.9605770E-01 |
| 0.0000E+00 | 9.0000000E-01 | 9.6837722E-01 | 9.9000000E-01 | 9.9683772E-01 |

While the method performs well for $c > 0.4$ with tables 1 and 2 taking less that 1.25 Min on a Precision Dell PC, the same cannot be said about $c < 0.4$. The computational time increases rapidly as $c$ approaches zero. An effort will be made to resolve this defect in the method.